\pdfoutput=1

\documentclass[preprint,prd,preprintnumbers,amssymb,superscriptaddress,nofootinbib]{revtex4}

\usepackage{epsfig,psfrag,graphics,verbatim}
\usepackage{graphicx}
\usepackage{dcolumn}
\usepackage{bm}
\usepackage{slashed}
\usepackage{color}
\usepackage{amsmath}
\usepackage[]{units}
\usepackage{ulem}

\usepackage{aas_macros}
\usepackage{hyperref}

\newcommand{\D}{\text{d}}

\begin{document}
\preprint{TUM-HEP 988/15}

\title{On the sensitivity of CTA to gamma-ray boxes\\ from multi-TeV dark matter}

\author{Alejandro Ibarra}
\affiliation{Physik-Department T30d, Technische Universit\"at M\"unchen, James-Franck-Stra\ss{}e, D-85748 Garching, Germany}

\author{Anna S.~Lamperstorfer}
\affiliation{Physik-Department T30d, Technische Universit\"at M\"unchen, James-Franck-Stra\ss{}e, D-85748 Garching, Germany}

\author{\\Sergio L\'opez-Gehler}
\affiliation{Physik-Department T30d, Technische Universit\"at M\"unchen, James-Franck-Stra\ss{}e, D-85748 Garching, Germany}
\affiliation{Excellence Cluster Universe, Technische Universit\"at M\"unchen, Boltzmannstra\ss{}e 2, D-85748, Garching, Germany}

\author{Miguel Pato}
\affiliation{The Oskar Klein Centre for Cosmoparticle Physics, Department of Physics, Stockholm University, AlbaNova, SE-106 91 Stockholm, Sweden}
\affiliation{Physik-Department T30d, Technische Universit\"at M\"unchen, James-Franck-Stra\ss{}e, D-85748 Garching, Germany}

\author{Gianfranco Bertone}
\affiliation{GRAPPA Institute, University of Amsterdam, Science Park 904, 1090 GL Amsterdam, The Netherlands}

\date{\today}

\begin{abstract}
Collider, direct and indirect searches for dark matter have typically little or no sensitivity to weakly interacting massive particles (WIMPs) with masses above a few TeV. This rather unexplored regime can however be probed through the search for distinctive gamma-ray spectral features produced by the annihilation of WIMPs at very high energies. Here we present a dedicated search for gamma-ray boxes -- sharp spectral features that cannot be mimicked by astrophysical sources -- with the upcoming Cherenkov Telescope Array (CTA). Using realistic projections for the instrument performance and detailed background modelling, a profile likelihood analysis is implemented to derive the expected upper limits and sensitivity reach after $100\,$h of observations towards a $2^\circ\times2^\circ$ region around the Galactic centre. Our results show that CTA will be able to probe gamma-ray boxes down to annihilation cross sections of $10^{-27}-10^{-26}\,\text{cm}^3\text{/s}$ up to tens of TeV. We also identify a number of concrete particle physics models providing thermal dark matter candidates that can be used as target benchmarks in future search campaigns. This constitutes a golden opportunity for CTA to either discover or rule out multi-TeV thermal dark matter in a corner of parameter space where all other experimental efforts are basically insensitive.
\end{abstract}

\maketitle

\section{Introduction}\label{sec:intro}
\par Among the myriad of dark matter candidates proposed, weakly interacting massive particles (WIMPs) still stand among the best motivated possibilities \cite{Bertone:2010zza,Bergstrom00,Jungman:1995df,Bertone05}. Under very plausible conditions, the freeze-out at very early times of the WIMP number density \cite{Zeldovich1,Zeldovich2,Chiu:1966kg} leads to a relic abundance which is in qualitative agreement with the observed dark matter abundance, $\Omega_{\rm dm} h^2=0.1198\pm 0.0015$ \cite{Planck:2015xua}, provided the thermally averaged annihilation cross section times the relative velocity of the dark matter particles at the time of freeze out is $\langle \sigma v\rangle \simeq 1\,{\rm pb}\cdot c$. The dark matter mass, however, cannot be predicted in this framework.

\par Most WIMP searches concentrate in the mass window between a few tens and a few hundreds of GeV, motivated by the possible relation of the dark matter problem with the stabilization of the electroweak scale against quadratically divergent quantum corrections. This is the case, for example, of the lightest neutralino in supersymmetric scenarios with R-parity conservation \cite{Ellis:1983ew} or the lightest Kaluza-Klein particle in scenarios with universal large extra dimensions \cite{Servant:2002aq}. However, on model independent grounds, the theoretically allowed window for WIMP masses is much wider, and could be as large as $\sim 118\,(84)\,$TeV for Majorana (Dirac) particles from the requirement of partial wave unitarity in the annihilation process \cite{Griest:1989wd} (using the latest measurement of the relic abundance \cite{Planck:2015xua}), or as low as $\sim 2\,$GeV, the renowned Lee-Weinberg limit for weakly interacting fermions \cite{Lee:1977ua}, or even lower for scalar dark matter \cite{Boehm:2003hm}. A complete investigation of the WIMP signals should therefore also include the high- and low-mass regions of the theoretically allowed parameter space.

\par In this work we focus on the search for heavy dark matter particles. This mass regime is especially challenging for the Large Hadron Collider (LHC), which will not be able to discover signals of WIMPs heavier than about $1.5\,$TeV after running at $14\,$TeV centre-of-mass energy with $300\,$fb$^{-1}$ of collected  data \cite{ATLAS:2013hta,CMS:2013xfa}. Future machines, such as a high-energy upgrade of the LHC with $33\,$TeV centre-of-mass energy \cite{Todesco:2011np} or a $100\,$TeV proton-proton collider \cite{100TeVBSM} will be able to extend the reach to higher masses, but not any time soon. The high-mass regime is also challenging for direct detection experiments, which reach their maximum sensitivity for WIMP masses of several tens of GeV. For instance, the sensitivity of LUX \cite{Akerib:2013tjd} (XENON100 \cite{Aprile:2012nq}) is best at masses of $\sim 33\,(55)\,$GeV under typical assumptions, and degrades quickly towards high masses. Nevertheless, the impressive limits on the scattering WIMP-nucleon cross section set by current experiments probe very heavy WIMPs in some concrete scenarios (see e.g.~\cite{Ibarra:2015fqa}). 

\par Indirect search experiments, and more specifically searches using gamma rays and neutrinos as messengers, may provide a very powerful probe for heavy dark matter. Some gamma-ray and neutrino telescopes are designed for the study of high-energy astrophysical phenomena and have their optimal sensitivity in the multi-TeV energy range, thus making them unique instruments to search for signals from the annihilation of heavy dark matter particles. There are several experiments currently operating which are probing WIMP masses above a few TeV, such as the gamma-ray instruments Fermi-LAT \cite{Ackermann:2012rg,Ackermann:2015zua}, H.E.S.S.~\cite{Abramowski:2011hc,Abramowski:2013ax}, MAGIC \cite{Aleksic:2013xea}  and VERITAS \cite{Aliu:2012ga}, as well as the neutrino telescopes Super-Kamiokande \cite{Tanaka:2011uf,Wendell:2014dka}, IceCube \cite{Aartsen:2012kia,Abbasi:2012ws}, ANTARES \cite{Hernandez-Rey:2014ssa} and Baksan \cite{Boliev:2013ai}. Unfortunately, the current limits are, for most scenarios, still far away from the required sensitivity to observe signals from heavy WIMP annihilations.

\par A notable exception is provided by scenarios generating gamma-ray boxes \cite{Ibarra:2012dw}. This sharp feature in the gamma-ray energy spectrum cannot be mimicked by any known astrophysical process, so the observation of a gamma-ray box would constitute an unambiguous signal for dark matter annihilations. Moreover, in contrast to annihilations generating gamma-ray lines \cite{Srednicki:1985sf,Rudaz:1986db,Bergstrom:1988fp} or internal bremsstrahlung \cite{Bergstrom:1989jr,Flores:1989ru}, which have cross sections suppressed with respect to the thermal value by a factor ${\cal O}(\alpha^2)$ and ${\cal O}(\alpha)$, respectively, with $\alpha$ the electromagnetic coupling constant, the gamma-ray box has an a priori unsuppressed cross section and can therefore generate an intense, and at the same time very characteristic, signal in the gamma-ray spectrum. In fact, in the most favourable scenarios, current instruments are sensitive to gamma-ray boxes generated by annihilations of WIMPs with masses as large as $\sim 10\,$TeV \cite{Ibarra:2013eda}. Motivated by the excellent sensitivity of current instruments to gamma-ray boxes, we investigate in this paper the prospects to observe this sharp feature with the future Cherenkov Telescope Array (CTA) \cite{ctasite,Consortium:2010bc}.

\par The paper is organised as follows. In Section \ref{sec:boxes} we review the most salient aspects of gamma-ray boxes and introduce a series of benchmark scenarios for our analysis. In Section \ref{sec:methodology} we describe our methodology to calculate the CTA reach and in Section \ref{sec:results} we present our results. Lastly, in Section \ref{sec:conclusions} we present our conclusions.

\section{Gamma-ray boxes from dark matter}\label{sec:boxes}
\par The search for gamma-ray boxes originated in dark matter annihilations or decays has been proposed and discussed at length elsewhere \cite{Ibarra:2012dw,Ibarra:2013eda}. Here, we simply recall the essential phenomenology to compute the gamma-ray flux at Earth and outline concrete dark matter models for illustration purposes.

\subsection{Phenomenology}\label{sec:boxpheno}
\par Gamma-ray boxes are the result of dark matter cascade annihilations or decays into (pseudo-)scalar intermediate particles that are short-lived and decay into photons. Consider for concreteness the case of Dirac dark matter particles $\chi$ annihilating into two scalars $\phi$ which decay into a pair of photons each: $\chi\bar{\chi}\to\phi\phi\to 4\gamma$ (for variations on this scenario we refer the reader to Refs.~\cite{Ibarra:2012dw,Ibarra:2013eda}). The scalars are sufficiently short-lived such that their decays occur basically in the same position as the dark matter annihilations -- in practice, since we are interested in regions of the Galaxy spanning hundreds of parsec, decay times below about $\unit[10^2-10^3]{yr}$ are sufficiently short for our purposes. In the rest frame of the parent scalar, the photons are monochromatic with energy $E_\gamma'=m_\phi/2$. After boosting to the lab frame, their energy $E_\gamma$ depends on the emission angle and lies in between the kinematic limits
\begin{equation}\label{eq:epm}
E_\pm=\left(1\pm \sqrt{ \delta_{\chi\phi} }\right)\frac{m_\chi}{2}  \, ,
\end{equation}
where the degeneracy parameter $\delta_{ij}=1-m_i^2/m_j^2$ has been used and $E_+$ ($E_-$) corresponds to a photon emitted in the forward (backward) direction. Since the intermediate state is a scalar, the angular distribution of the photons is isotropic (in the rest frame of $\phi$) and all lab-frame energies between $E_+$ and $E_-$ are equally populated, giving rise to a photon spectrum with a characteristic box shape:
\begin{equation}
\frac{\D N_\gamma}{\D E_\gamma}=\frac{4}{\Delta E}\,\Theta(E-E_-)\Theta(E_+-E) \, .
\end{equation}
In the formula above, $\Delta E \equiv E_+-E_-$ is the width of the box and $\Theta(x)$ is the Heaviside step function. The photon spectrum at Earth induced by such injection spectrum reads
\begin{equation}\label{eq:unconvolphi}
\frac{\D \Phi_\text{dm}}{\D E_\gamma} = \frac{\langle \sigma v \rangle_{0}^{4\gamma}}{16\pi m_\chi^2} \, \frac{\D N_{\gamma}}{\D E_{\gamma}} \, J_{\text{ann}} \qquad \text{with} \qquad J_{\text{ann}} = \int_{\Delta \Omega}{\D\Omega \, \int_{\text{los}}{\D s \, \rho_{\text{dm}}^2 } }
\end{equation}
for symmetric Dirac dark matter particles, where $\langle \sigma v \rangle_{0}^{4\gamma}$ is the present-day thermally averaged annihilation cross section of the cascade $\chi\bar{\chi}\to\phi\phi\to 4\gamma$, $J_{\text{ann}}$ represents the usual J-factor, $\Delta\Omega$ is the observed field of view, $s$ parameterizes the direction down the line of sight defined by Galactic coordinates $(\ell,b)$ and $\rho_{\text{dm}}$ is the dark matter density. In this work, we shall focus on a $2^\circ \times 2^\circ$ field of view around the Galactic centre (i.e.~$|\ell|\leq1^\circ$ and $|b|\leq1^\circ$) and consider different dark matter profiles: Einasto \cite{Navarro:2003ew,Merritt2006} with scale radius $r_{s}=\unit[20]{kpc}$ and shape parameter $\alpha=0.17$, Navarro-Frenk-White (NFW) \cite{NFW1996,Navarro:1996gj} with $r_{s}=\unit[20]{kpc}$, a cuspy generalised $(\alpha,\beta,\gamma)$ profile \cite{1990ApJ...356..359H,Merritt2006} with $r_{s}=\unit[20]{kpc}$ and $(\alpha,\beta,\gamma)=(1,3,1.2)$ (motivated by the gamma-ray excess towards the Galactic centre \cite{Berlin:2014pya}) and a cored generalised $(\alpha,\beta,\gamma)$ profile with $r_{s}\simeq\unit[4.4]{kpc}$ and $(\alpha,\beta,\gamma)\simeq(2.9,2.5,0)$ (motivated by recent high-resolution simulations including baryons \cite{Mollitor:2014ara}; we use halo B specified in Tab.~3 in that reference). All profiles are normalised to $\rho_0=\unit[0.4]{GeV/cm^3}$ at the position of the Sun, adopting a distance to the Galactic centre $R_0=8.5\,$kpc. For reference, the J-factors towards our $2^\circ \times 2^\circ$ target region read $J_\text{ann}=\unit[1.02 \times 10^{22}, 6.34 \times 10^{21}, 3.63 \times 10^{22}, 1.60 \times 10^{20}]{GeV^2sr/cm^5}$ for Einasto, NFW, cuspy and cored profiles, respectively.

\par The flux in Eq.~\eqref{eq:unconvolphi} assumes no attenuation of the photons in their route to Earth. However, as pointed out right after the discovery of the cosmic microwave background (CMB) \cite{1966PhRvL..16..252G,1967PhRv..155.1408G,1966PhRvL..16..479J}, the universe is not transparent to high-energy gamma rays, their mean free path due to pair production on background photons reaching a minimum of $\sim8\,$kpc for $E_\gamma\simeq2000\,$TeV (e.g.~\cite{DeAngelis:2013jna}). Since our region of interest is the centre of the Milky Way, at around $8\,$kpc from us, and we are in fact interested in energies up to hundreds of TeV, it is pertinent to calculate the optical depth of gamma rays due to pair production on the interstellar radiation field. The latter is composed of starlight (mostly ultraviolet), dust-reprocessed emission (mainly infrared) and the CMB. The attenuation due to starlight and dust emission becomes relevant at photon energies of a few tens of TeV, while the influence of CMB kicks in at around $\unit[200]{TeV}$. We compute the optical depth $\tau$ due to pair production following standard techniques (see e.g.~\cite{1996ApJ...456..124M}) and using the interstellar radiation field as found in \cite{Porter:2005qx,Moskalenko:2005ng} at Galactocentric radii $R=0,\,4,\,12$ and $\unit[16]{kpc}$. Taking account of photon attenuation makes the J-factor depend on energy:
\begin{equation}\label{eq:Jtau}
J_{\text{ann}}^\tau(E_\gamma) = \int_{\Delta \Omega}{\D \Omega \, \int_{\text{los}}{\D s \, e^{-\tau(E_\gamma,s,\Omega)} \rho_{\text{dm}}^2 } } \, ,
\end{equation}
where the optical depth $\tau(E_\gamma,s,\Omega)$ is computed along the line of sight. For our $2^\circ \times 2^\circ$ region of interest and the Einasto profile, we obtain $J_{\text{ann}}^\tau/J_{\text{ann}}\simeq 1,\,0.99,\,0.75,\,0.35$ for $E_\gamma=\unit[10]{GeV},\,\unit[1]{TeV},\,\unit[100]{TeV},\,\unit[1000]{TeV}$, respectively.

\subsection{Concrete models}\label{sec:boxmodels}
\par Physical models that feature boxed-shaped spectra are usually based on the breaking of global symmetries and involve Dirac dark matter \cite{Lee:2012bq, Lee:2012wz, Ibarra:2013eda, Nomura:2008ru, Mardon:2009rc, Mardon:2009gw} (see also \cite{Garcia-Cely:2013wda} for a model with chiral fermion dark matter). We first discuss the setup developed in Refs.~\cite{Lee:2012bq, Lee:2012wz, Ibarra:2013eda}, where the Peccei-Quinn mechanism \cite{PhysRevLett.38.1440} is invoked and the resulting pseudoscalars play the role of the intermediate states $\phi$ referred to in the previous subsection. The Lagrangian $\mathcal{L}$ of this model reads
\begin{equation}
\mathcal{L}=\mathcal{L}_\text{SM}+\mathcal{L}_\chi+\mathcal{L}_S+\mathcal{L}_\text{int} \, ,
\end{equation}    
where $\mathcal{L}_\text{SM}$ is the Standard Model Lagrangian, which includes the potential for the Higgs field, whereas
\begin{align}
\mathcal{L}_\chi &=i\bar{\chi}\gamma^\mu \partial_\mu \chi \;,  \\
\mathcal{L}_S &=\partial_\mu S \partial^\mu S^*-m_S^2|S|^2-\lambda_S |S|^4+
\left( \frac{1}{2} m_S^{\prime\,2} S^2 + \text{c.c.}\right) \, ,\\
\mathcal{L}_\text{int}&=-\lambda_\chi\left(S\bar{\chi}P_L\chi +S^*\bar{\chi}P_R\chi\right) -2\lambda_{H,S}|H|^2|S|^2 \label{eq:L_int}
\end{align}
are, respectively, the parts of the Lagrangian involving only the field $\chi$, only the field $S$ and the interaction Lagrangian. We further assume that the field $S$ acquires a  vacuum expectation value, so that the field $S$ can be decomposed as $S=\langle S\rangle +\frac{1}{\sqrt{2}}(s+ia)$. The interaction Lagrangian in Eq.~(\ref{eq:L_int}) then generates a mass term for the Dirac field $\chi$, as well as a mixing term between the CP-even scalar $s$ and the Standard Model Higgs boson. This mixing term, in particular, permits the thermalisation of all new states with the Standard Model plasma in the early universe.

\par If the mass term $m_S^{\prime\,2}$ vanishes, the Lagrangian displays a global $U(1)_\text{PQ}$  symmetry under the field transformations $\chi \to e^{i \gamma_5 \alpha}\chi $ and $S \to e^{2i\alpha}S$, while all the Standard Model fields remain invariant. This symmetry is broken in the vacuum, however a residual $Z_2$ symmetry remains, under which $\chi$ is odd and all other fields are even. The exact $Z_2$ symmetry then guarantees the stability of $\chi$, which constitutes our dark matter candidate. The observed dark matter abundance, moreover, can be explained with the freeze-out of $\chi$ from the thermal plasma due to the annihilations $\chi\bar{\chi} \to aa,\,ss,\,sa$, when kinematically allowed. We note that the field $a$, being the Goldstone boson resulting from the spontaneous breaking of the global $U(1)_\text{PQ}$ symmetry, remains massless and is therefore also absolutely stable, thus constituting a candidate for dark radiation \cite{Weinberg:2013kea,Garcia-Cely:2013nin}.

\par Here, we instead allow the mass term $m_S^{\prime\,2}$ to be non-vanishing, thus  breaking explicitly the $U(1)_\text{PQ}$ symmetry. This variation does not affect the stability of the dark matter particle, however it introduces a mass for the pseudoscalar, $m_a=m_S^{\prime}$, which can then decay into Standard Model particles. In particular, the pseudoscalar could decay into two electroweak gauge bosons if the model is extended with new heavy fermions with electroweak charges that couple to the complex scalar $S$, thus generating the following term in the interaction Lagrangian via the anomaly loop:
\begin{equation}
\mathcal{L}_\text{int} \supset \sum_{i=1,2}\frac{c_i\alpha_i}{8\pi v_s}a\,F_{\mu\nu}^i\tilde{F}^{i\,\mu\nu} \, ,
\end{equation}
and which induces the decays $a \to \gamma\gamma,\,\gamma Z,\,ZZ,\,W^+W^-$. Here, $c_i$ are constant parameters which depend on the assigned charges of heavy fermions in the anomaly loop, the parameters $\alpha_i$ are given by $\alpha_i=g_i^2/4\pi$ where $g_{1,2}$ are the coupling constants of $U(1)_\text{Y}$ and $SU(2)_\text{L}$, respectively, and finally $\tilde{F}_{\mu\nu}=\epsilon_{\mu\nu\rho\sigma}F^{\rho\sigma}/2$ is the dual field strength tensor. 

\par The interactions described by the Lagrangian in Eq.~(\ref{eq:L_int}) induce the dark matter annihilations $\chi\bar{\chi} \to aa,\,ss,\,sa$. While $s$ decays producing a continuum of gamma-rays, through  the mixing with the Higgs boson, the pseudoscalar $a$ can decay into one or two photons, hence producing a gamma-ray box. In order to assess the prospects to observe signals of this model we consider three different benchmark scenarios taken from Ref.~\cite{Ibarra:2013eda}, which assume $m_s=m_a$ and that we label A1, A2 and A3, corresponding to narrow, intermediate and wide boxes, respectively (see Tab.~\ref{tab:annbr}). In all three cases the ratios for $\chi\bar{\chi} \to aa$ and $\chi\bar{\chi} \to as$, needed to produce a box spectrum, sum up to $50\%$ or more. The decay of $a$ into vector bosons depends on the constants $c_i$; we adopt $c_1/c_2=3$ for which the branching ratio BR$(a\to \gamma\gamma)$ is $100\%$ for $m_a\lesssim 2m_W$ and gradually decreases to $20\%$ at high pseudoscalar masses (for exact expressions see Ref.~\cite{Ibarra:2013eda}). With these figures, the annihilation cross section in Eq.~\eqref{eq:unconvolphi} reads
\begin{equation}\label{eq:sigmaBRA}
\langle \sigma v \rangle_{0}^{4\gamma} = \langle \sigma v \rangle_{0} \left( \text{BR}(\chi\bar{\chi} \to aa) + \frac{\text{BR}(\chi\bar{\chi} \to as)}{2} \right) \text{BR}(a\to \gamma\gamma) \, ,
\end{equation}
where $\langle \sigma v \rangle_{0}$ is the present-day total annihilation cross section. The ratio of $\langle \sigma v \rangle_{0}$ to the corresponding value at thermal decoupling  $\langle \sigma v \rangle_{\text{th}}$ is shown in Tab.~\ref{tab:annbr}. We compute the thermal average in the non-relativistic limit as in Ref.~\cite{Ibarra:2013eda}; we have however verified that a full treatment \cite{1991NuPhB.360..145G,1997PhRvD..56.1879E} leads to changes of less than 10\% on the ratio $\langle \sigma v \rangle_{0}/\langle \sigma v \rangle_{\text{th}}$. In order to test thermal candidates with the observed relic abundance, we shall take the canonical value $\langle \sigma v \rangle_{\text{th}}=\unit[6\times 10^{-26}]{cm^3/s}$. The factor $2$ with respect to the usual $\unit[3\times 10^{-26}]{cm^3/s}$ is because the models considered here involve Dirac dark matter, not Majorana.

\par Additionally, we analyse the framework presented in \cite{Nomura:2008ru, Mardon:2009rc, Mardon:2009gw}, designed to interpret the positron excesses observed by PAMELA \cite{Adriani:2008zr}, Fermi-LAT \cite{FermiLAT:2011ab} and AMS-02 \cite{Aguilar:2013qda} while circumventing the non-observation of large gamma-ray fluxes due to inverse Compton scattering \cite{Mardon:2009rc} or a large antiproton flux. In these models, dark matter couples to a scalar $s$ and a pseudoscalar $a$ as in the previous framework, but now the s-wave annihilation channel $\chi\bar{\chi}\to a s$ is dominant \cite{Nomura:2008ru}. Therefore, BR$(\chi\bar{\chi}\to a s)\simeq 1$ and $\langle\sigma v\rangle_{0}=\langle\sigma v\rangle_{\text{th}}$ as indicated in Tab.~\ref{tab:annbr} for the benchmark scenario labelled B. Again, we take the canonical value $\langle \sigma v \rangle_{\text{th}}=\unit[6\times 10^{-26}]{cm^3s^{-1}}$. Notice that this framework aims at explaining experimental data and therefore the masses of the involved particles are restricted to the ranges $m_\chi\simeq\mathcal{O}(\text{TeV})$, $m_s\ll m_\chi$ and $m_{a}=\unit[360-800]{MeV}$, giving rise to wide boxes only. For definiteness, we assume $m_a/m_\chi=m_s/m_\chi=0.1$ throughout (note that the phenomenology of wide boxes is insensitive to the exact mass ratio, see \cite{Ibarra:2012dw,Ibarra:2013eda}). Furthermore, the decay of $a$ is primarily leptonic leading to a branching ratio BR$(a\to\gamma\gamma)\simeq 10^{-3}$, which is however compensated by a Sommerfeld enhancement\footnote{In principle, also scenarios A give rise to Sommerfeld enhancement, but to the best of our knowledge no previous study has addressed this behaviour. A full treatment lies outside of the scope of the current paper and is left for future work.} $B_S\sim 10^3$ on the present-day annihilation rate \cite{Nomura:2008ru}. In the end the annihilation cross section entering Eq.~\eqref{eq:unconvolphi} is
\begin{equation}\label{eq:sigmaBRB}
\langle \sigma v \rangle_{0}^{4\gamma} = B_S \, \langle \sigma v \rangle_{0} \frac{\text{BR}(\chi\bar{\chi} \to as)}{2} \text{BR}(a\to \gamma\gamma) \, .
\end{equation}
\par The two classes of models presented above will serve as benchmarks for the following discussions, where Eqs.~\eqref{eq:sigmaBRA} and \eqref{eq:sigmaBRB} (cf.~also Tab.~\ref{tab:annbr}) will be used as target velocity-averaged cross sections to test the performance of the CTA.

\begin{table}
\centering
\begin{tabular}{l|c|c|c|c|c|c}
\hline
\hline
 & $m_a/m_\chi$ & BR$(\chi\bar{\chi}\to as)$ &BR$(\chi\bar{\chi}\to aa)$ &BR$(\chi\bar{\chi}\to ss)$ & $\langle \sigma v \rangle_{0}/\langle \sigma v \rangle_{\text{th}}$ & References \\
\hline
A1 (narrow) & $0.999$ & $0.99$&$2\times 10^{-3}$&$4\times 10^{-3}$ & $0.13$ & \cite{Lee:2012bq, Lee:2012wz, Ibarra:2013eda}\\
\hline
A2 (intermediate) & $0.9$ &  $0.64$&$0.12$&$0.24$ & $0.76$ & \cite{Lee:2012bq, Lee:2012wz, Ibarra:2013eda}\\
\hline
A3 (wide) & $0.1$ &  $0.25$&$0.25$&$0.50$ & $0.96$ & \cite{Lee:2012bq, Lee:2012wz, Ibarra:2013eda}\\
\hline
\hline

B \hspace{0.09cm} (wide) & $0.1$ & 1 & 0 & 0 & 1 & \cite{Nomura:2008ru, Mardon:2009rc, Mardon:2009gw}\\

\hline
\hline
\end{tabular}
\caption{\small Details of the concrete models based on Refs.~\cite{Lee:2012bq, Lee:2012wz, Ibarra:2013eda} (A1, A2, A3) giving rise to narrow, intermediate and wide boxes and on Refs.~\cite{Nomura:2008ru, Mardon:2009rc, Mardon:2009gw} (B) where only wide boxes are produced. In all cases we have assumed $m_s=m_a$.}
\label{tab:annbr}
\end{table}

\section{Methodology}\label{sec:methodology}
\par In order to derive the prospects for CTA to detect or rule out gamma-ray boxes we need a detailed account of the instrument properties, the expected background and a statistical procedure to convey our results. These three topics are now discussed in turn.

\subsection{CTA performance}\label{sec:performance}
\par The upcoming Cherenkov Telescope Array (CTA) \cite{ctasite,Consortium:2010bc} will consist of two sites, one in each hemisphere, equipped with three different kinds of telescopes in order to cover a wide energy range from $\unit[10]{GeV}$ to about $\unit[100]{TeV}$. CTA is expected to outperform existing arrays by one order of magnitude \cite{Consortium:2010bc}, reaching differential sensitivities down to $\unit[\text{a few}]{milliCrab}$ at $\unit[1]{TeV}$ after $\unit[50]{h}$ of observations \cite{Vandenbroucke:2011sh,Bernlohr:2012we}. Detailed information on Monte Carlo design studies can be found in Ref.~\cite{Bernlohr:2012we}. In our work we implement the performance of the southern candidate array I, which is a well-studied, balanced configuration consisting of 3 large, 18 medium and 56 small size telescopes. Above a few TeV the effective area of this array exceeds $\unit[10^6]{m^2}$ and the energy resolution is better than 10\%. We use throughout the detailed effective area $A_\text{eff}$ and energy resolution $\sigma_E$ as found using the baseline MPIK analysis (cf.~Figs.~15 and 18 in Ref.~\cite{Bernlohr:2012we}). Note that the effective area in Ref.~\cite{Bernlohr:2012we} is given at the analysis level so that we take a photon acceptance $\epsilon_\gamma=1$. Our study does not explore the angular resolution of the instrument, but instead focuses on a $2^\circ \times 2^\circ$ region around the Galactic centre (as mentioned in Sec.~\ref{sec:boxpheno}) for which an observation time $\Delta t = \unit[100]{h}$ is reasonable given the high astrophysical interest of this region. We further assume an electron acceptance $\epsilon_e=1$ and a proton acceptance of $\epsilon_p=0.01-0.20$ (see Sec.~\ref{sec:backgrounds} for a full expression). Finally, let us point out that recently an updated study of the expected performance of CTA has been made available \cite{ctasiteperformance}; we also implement this updated performance for which effective area and energy resolution are better than those of Ref.~\cite{Bernlohr:2012we} and the proton acceptance reads instead $\epsilon_p=0.001-0.02$ (again, see Sec.~\ref{sec:backgrounds} for further details). The performance based on Ref.~\cite{Bernlohr:2012we} is used throughout the work and we comment on the results obtained with the updated performance \cite{ctasiteperformance} in Sec.~\ref{sec:results} (cf.~in particular Figs.~\ref{fig:limits} and \ref{fig:HessFermi}).

\subsection{Backgrounds}\label{sec:backgrounds}
\par The search for dark matter with imaging air Cherenkov telescopes is usually hindered by three sorts of background:
\begin{itemize} 
\item First, hadronic cosmic rays (mainly protons) initiate showers whose Cherenkov radiation may be misread as coming from a photon-initiated shower. It is actually possible to discriminate hadronic and electromagnetic cascades -- typical hadron acceptances vary between 1\% and 20\% for a 70\% gamma-ray acceptance \cite{2004NIMPA.516..511B}. However, cosmic rays are still a sizeable background since they largely outnumber photons. 
\item Second, showers induced by electrons or positrons are indistinguishable from those triggered by photons at the same energy (apart from a slightly later shower maximum on average for photons \cite{Aharonian:2008aa}) and constitute an important background as well. 
\item Third, photons from astrophysical or other origin ultimately limit the sensitivity of dark matter searches and can only be mitigated with a wise choice of the target field of view and analysis technique. 
\end{itemize} 

In the following we describe in detail the parameterizations used in our analysis for each type of background.  Let us start with cosmic-ray protons. We follow Ref.~\cite{Hoerandel:2002yg} where it was found that the power law
\begin{equation}
\frac{\D^2 \Phi_{p}}{\D E_p \D\Omega}= 8.73 \times 10^{-6} \left(  \frac{E_p}{\text{TeV}}\right)^{-2.71} \, \unit{TeV^{-1}cm^{-2}s^{-1}sr^{-1}}
\end{equation}
provides a good fit to a wide array of data prior to 2002 from around $10\,$GeV up to knee energies. Including the latest proton data \cite{Maurin:2013lwa} changes only slightly the best-fit power law, so we use the above parameterization throughout. Notice that for primary energies $E_0\gtrsim \unit[100]{GeV}$ the Cherenkov yield of a photon shower is a factor $\rho\simeq2-3$ larger than that of a proton shower of the same energy, and both yields are roughly proportional to $E_0$ \cite{1997JPhG...23.1013F}. Therefore, a shower triggered by a proton with energy $E_p$  has the same yield as a photon-initiated shower of a smaller primary energy $E_\gamma\simeq E_p/\rho$. Actually, there are several shower parameters other than the Cherenkov yield used to discriminate photons from hadrons \cite{1997JPhG...23.1013F}, but, in the absence of detailed information about the typical energies of misreconstructed protons, we use the relation above with $\rho=3$ in computing the background flux due to protons.

\par The all-electron spectrum has also been precisely determined, namely by AMS-02 \cite{Aguilar:2014fea} below several hundred GeV and by H.E.S.S.~\cite{Aharonian:2008aa} at TeV energies. AMS-02 data follows essentially a power law
\begin{equation}
\frac{\D^2 \Phi_{e}^{\text{le}}}{\D E_e \D\Omega}= 9.93 \times 10^{-9} \left( \frac{E_e}{\text{TeV}}\right)^{-3.17} \, \unit{TeV^{-1}cm^{-2}s^{-1}sr^{-1}}
\end{equation}
for energies above $\unit[30.2]{GeV}$, while H.E.S.S.~data yields
\begin{equation}
\frac{\D^2 \Phi_{e}^{\text{he}}}{\D E_e \D\Omega}= 1.17 \times 10^{-8} \left(  \frac{E_e}{\text{TeV}}\right)^{-3.9} \, \unit{TeV^{-1}cm^{-2}s^{-1}sr^{-1}} \, .
\end{equation}
In order to ensure a smooth transition between the two regimes the total all-electron flux reads $\D^2 \Phi_{e}/\D E_e \D\Omega = [ (\D^2 \Phi_{e}^{\text{le}}/\D E_e \D\Omega)^{-2} + (\D^2 \Phi_{e}^{\text{he}}/\D E_e \D\Omega)^{-2} ]^{-1/2}$.

\par Finally, the background due to gamma rays of astrophysical origin depends heavily on the chosen target field of view. In our $2^\circ \times 2^\circ$ region of interest there are two main gamma-ray components at high energies. On the one hand, the Galactic ridge emission spanning the region $|\ell|<0.8^\circ$ and $|b|<0.3^\circ$ has been detected by H.E.S.S.~\cite{Aharonian:2006au} and is well described by the power law
\begin{equation}
\frac{\D^2 \Phi_{\gamma,\text{gr}}}{\D E_\gamma \D \Omega}= 1.73 \times 10^{-8} \left(  \frac{E_\gamma}{\text{TeV}}\right)^{-2.29} \, \unit{TeV^{-1}cm^{-2}s^{-1}sr^{-1}} \, .
\end{equation}
On the other hand, data from Fermi-LAT and H.E.S.S.~show a point source coincident with the Galactic centre (to within the resolution of the instruments) whose energy spectrum in the range $\unit[5-100]{GeV}$ is well fitted by \cite{Chernyakova:2011zz} 
\begin{equation}
\frac{\D \Phi_{\gamma,\text{gc}}^{\text{le}}}{\D E_\gamma}= 1.11 \times 10^{-12} \left( \frac{E_\gamma}{\text{TeV}}\right)^{-2.68} \, \unit{TeV^{-1}cm^{-2}s^{-1}}
\end{equation}
and in the range $\unit[160]{GeV}-\unit[30]{TeV}$ by \cite{Aharonian:2006wh}
\begin{equation}
\frac{\D \Phi_{\gamma,\text{gc}}^{\text{he}}}{\D E_\gamma}= 2.34 \times 10^{-12} \left(  \frac{E_\gamma}{\text{TeV}}\right)^{-2.25} \, \unit{TeV^{-1}cm^{-2}s^{-1}} \,\, .
\end{equation}
A smooth transition between low and high energies is achieved using $\D \Phi_{\gamma,\text{gc}}/\D E_\gamma = [ (\D \Phi_{\gamma,\text{gc}}^{\text{le}}/\D E_\gamma)^5 + (\D \Phi_{\gamma,\text{gc}}^{\text{he}}/\D E_\gamma)^5]^{1/5}$.

\par The overall background flux then reads
\begin{equation}\label{eq:backg}
\frac{\D \Phi_{\text{bkg}}}{\D E_\gamma} \left(E_\gamma \right) = 
\Delta \Omega_{2\times2} \left(
 \epsilon_p(E_\gamma)  \rho \frac{\D^2 \Phi_{p}}{\D E_p \D\Omega}\left(\rho E_\gamma\right) +
 \epsilon_e \frac{\D^2 \Phi_{e}}{\D E_e \D\Omega} \left(E_\gamma \right) +
 \epsilon_\gamma\frac{\D^2 \Phi_{\gamma,\text{gr}}}{\D E_\gamma \D \Omega} \left(E_\gamma \right)  
\right) +
\epsilon_\gamma \frac{\D \Phi_{\gamma,\text{gc}}}{\D E_\gamma} \left(E_\gamma \right) \, ,
\end{equation}
where $\Delta \Omega_{2\times2}=1.22\times 10^{-3}\,\text{sr}$ is the solid angle of our $2^\circ \times 2^\circ$ region of interest. Notice that, in order to be as conservative as possible, we upscale the flux found in the Galactic ridge ($|\ell|<0.8^\circ$ and $|b|<0.3^\circ$) to match our target field of view assuming the same output per unit solid angle. Furthermore, we have checked that our cosmic-ray background rate matches the one reported in Ref.~\cite{Bernlohr:2012we} (see Fig.~16 therein) within a factor $1.5-3$ if we take $\epsilon_p(E_\gamma)=0.01+0.02(E_\gamma/\unit[20]{TeV})^{1.4}$, i.e.~a benchmark 1\% proton acceptance below $\sim 1\,$TeV that grows up to around 20\% at $100\,$TeV. This is roughly in line with Ref.~\cite{Lefranc:2015pza}, where a full CTA simulation is used to set the cosmic-ray background rate. We repeated the same procedure with the updated CTA performance \cite{ctasiteperformance}, obtaining $\epsilon_p(E_\gamma)=0.001+0.002(E_\gamma/\unit[20]{TeV})^{1.4}$.

\par
\begin{figure}[tp]
\begin{center}
\includegraphics[width=.49\textwidth]{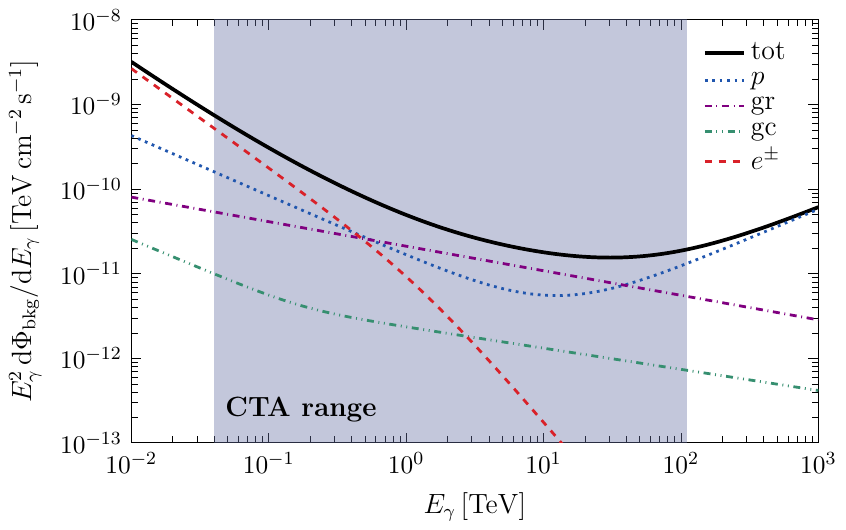}
\includegraphics[width=.49\textwidth]{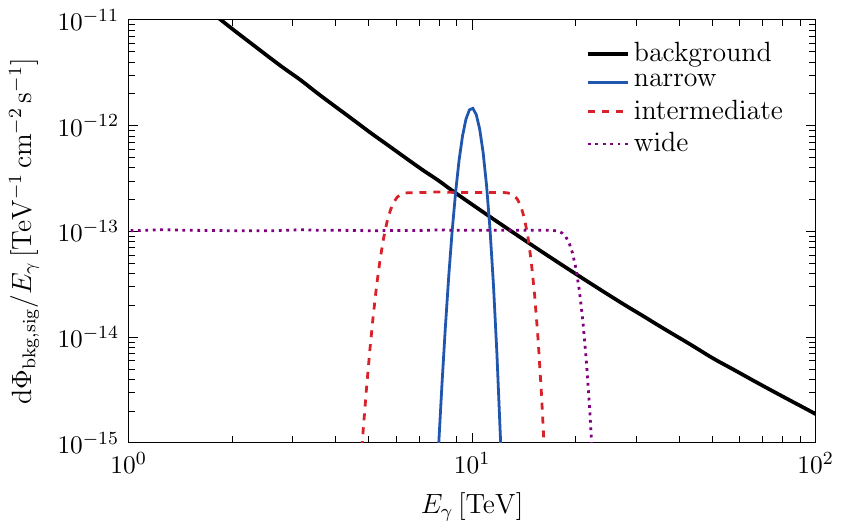}
\caption{\small The background for CTA gamma-ray searches in a $2^\circ \times 2^\circ$ region around the Galactic centre. The left panel shows the breakdown of the total background in Eq.~\eqref{eq:backg} into protons, electrons and positrons, and gamma rays from both the Galactic centre source and the Galactic ridge region. The right panel compares the total background flux with narrow, intermediate and wide gamma-ray boxes induced by a dark matter scenario with an Einasto profile, $\langle \sigma v \rangle_{0}^{4\gamma}=\unit[10^{-24}]{cm^3/s}$, $m_\chi=\unit[20]{TeV}$ and $m_\phi/m_\chi=0.999,\,0.9,\,0.1$, respectively. Both background and signal in the right panel are convoluted with the CTA energy resolution described in Sec.~\ref{sec:performance}. Here we use the CTA performance of Ref.~\cite{Bernlohr:2012we}.
}
\label{fig:bkg}
\end{center}
\end{figure}

 The breakdown of the background in Eq.~\eqref{eq:backg} is shown in the left plot of Fig.~\ref{fig:bkg}. Clearly, below a few hundred GeV cosmic-ray electrons and positrons represent the largest background, whereas at TeV energies and above the diffuse emission from the Galactic ridge and eventually protons overshadow all other components. The right panel of Fig.~\ref{fig:bkg} shows a side-by-side comparison of the overall background  and the signal from Eq.~\eqref{eq:unconvolphi} for the benchmark values $\langle \sigma v \rangle_{0}^{4\gamma}=\unit[10^{-24}]{cm^3/s}$, $m_\chi=\unit[20]{TeV}$ and $m_\phi/m_\chi=0.999,\,0.9,\,0.1$.

\subsection{Limits and sensitivity}\label{sec:likelihood}
\par After outlining the properties of CTA and defining background and signal fluxes, we can now generate mock data for CTA. For definiteness, $N_b=200$ energy bins per decade are set up in the energy range $E_\gamma\simeq\unit[40]{GeV}-\unit[110]{TeV}$. Assuming a generic flux $\D \Phi_{\text{tot}}/\D E_\gamma$, the expected number of counts $n_\text{exp}^i$ in bin $i$ of width $\Delta E_i$ is simply
\begin{equation}\label{eq:nexp} 
n_\text{exp}^i = \Delta t \int_{\Delta E_i} \D E \int \D E'\, R(E,E')\, A_\text{eff}(E') \, \frac{\D \Phi_{\text{tot}}}{\D E_\gamma} (E') \, ,
\end{equation}
where $R(E,E')$ is the energy resolution of the instrument, assumed Gaussian with standard deviation given by $\sigma_E$ (cf.~Sec.~\ref{sec:performance}). The mock number of counts $n_\text{obs}^i$ is drawn from a Poissonian distribution with mean $n_\text{exp}^i$.

\par Given a data set $\{n_\text{obs}^{i=1,...,N}\}$, we make use of a profile likelihood analysis \cite{Rolke:2004mj,VertongenWeniger,Bringmann:2011ye,Bringmannetal,Weniger:2012tx} in order to search for gamma-ray boxes with fixed mass configuration $(m_\chi,m_\phi)$. Our flux model $\D \Phi_{\text{tot}}/\D E_\gamma$ includes a power-law background (with normalisation $A$ and spectral index $-\gamma$) and a putative gamma-ray box (with strength set by $S\equiv\langle \sigma v \rangle_{0}^{4\gamma}$), in a total of three parameters. For each bin $i$ in the data set, the expected number of counts $n_{\text{exp}}^i(A,\gamma,S)$ follows from Eq.~\eqref{eq:nexp}, and the likelihood is written as a product of Poissonian probabilities:
\begin{equation}\label{eq:Lpoisson}
\mathcal{L}= \prod_i{  \frac{ (n_{\text{exp}}^i)^{n_{\text{obs}}^i} \, \exp(-n_{\text{exp}}^i) } {n_{\text{obs}}^i!} } \, .
\end{equation}
We use the sliding window technique \cite{Pullen:2006sy,Abdo:2010nc,VertongenWeniger,Bringmann:2011ye,Bringmannetal,Weniger:2012tx} such that the product in the expression above is restricted to the bins inside a given energy window $[\bar{E}/\sqrt{\epsilon},\bar{E}\sqrt{\epsilon}]$.  The quantity $-2\ln\mathcal{L}$ is then minimised over $(A,\gamma)$ in order to define the profile likelihood $-2\ln\mathcal{L}_{\text{prof}}$ as a function of $S$, and over $(A,\gamma,S)$ to find the global minimum (i.e.~best fit) $-2\ln\mathcal{L}_{\text{bf}}$ attained at $(A_\text{bf},\gamma_\text{bf},S_\text{bf})$. The best fit under the null hypothesis follows immediately from the profile likelihood as $-2\ln\mathcal{L}_{\text{bf}}^0=-2\ln\mathcal{L}_{\text{prof}}(S=0)$.

\par The aim of this paper is to infer the prospects for CTA in the search for gamma-ray boxes, namely upper limits in case of non-observation and sensitivity reach otherwise. For the upper limits, the starting point is to generate 300 sets of mock data assuming background only, i.e.~with $\D \Phi_{\text{tot}}/\D E_\gamma=\D \Phi_{\text{bkg}}/\D E_\gamma$ in Eq.~\eqref{eq:nexp}. Next, for each data set and any given combination $(m_\chi,m_\phi)$, we compute the profile likelihood and find the one-sided 95\% confidence level (CL) upper limit $S_\text{ul}>S_\text{bf}$ defined as $-2\ln\mathcal{L}_\text{prof}(S_\text{ul})=-2\ln\mathcal{L}_\text{bf}+2.71$ \cite{1997sda..book.....C}. For the sensitivity reach, having chosen a specific mass configuration $(m_\chi,m_\phi)$, we start by drawing 300 realisations of mock data assuming signal plus background, i.e.~$\D \Phi_{\text{tot}}/\D E_\gamma=\epsilon_\gamma\D \Phi_{\text{dm}}/\D E_\gamma + \D \Phi_{\text{bkg}}/\D E_\gamma$, where the signal is computed with a relatively small value of $S$ (i.e.~$\langle \sigma v\rangle_{0}^{4\gamma}$). We then determine the average of the test statistic $TS=-2\left( \ln\mathcal{L}_\text{bf}^0 - \ln\mathcal{L}_\text{bf} \right)$ over the 300 realisations and repeat the procedure with a higher value of $S$. The $5\sigma$ sensitivity reach $S_{5\sigma}$ corresponds to the smallest annihilation cross section for which the average $TS$ is larger than 23.7 (see e.g.~\cite{VertongenWeniger}).

\par Let us finally comment on the energy windows used in our analysis. We set the central energy of all windows to $\bar{E}=E_+$ (cf.~Eq.~\eqref{eq:epm}) and adopt three different widths $\epsilon$: $\epsilon_{1.2}=1.2$, $\epsilon_{1.5}=1.5$ and $\epsilon_2=2$. These windows are chosen to test the robustness of our results and, at the same time, ensure that the background inside the window is well fitted by a power law. Specifically, we started by taking the 300 mock data sets generated assuming background only (see above), and fitted each data set to a pure power law using windows of constant widths (spanning the range $\epsilon=1.2-10$) and centred across the full CTA energy range. For each window, we thus obtain 300 $\chi^2$ values for the best fit power laws and test whether this distribution is plausibly drawn from a $\chi^2$ distribution with the appropriate degrees of freedom (number of bins minus two fitted parameters). Windows for which the p-value of this distribution test is smaller than 0.01 are rejected. The window of width $\epsilon_2$ is in particular the largest window surviving this criterion, so we shall adopt it as our default window throughout the work, while still using $\epsilon_{1.2}$ and $\epsilon_{1.5}$ for completeness.

\section{Results}\label{sec:results}

\begin{figure}[tp!]
\begin{center}
\hspace{-.25\textwidth}
\includegraphics[height=.4\textwidth]{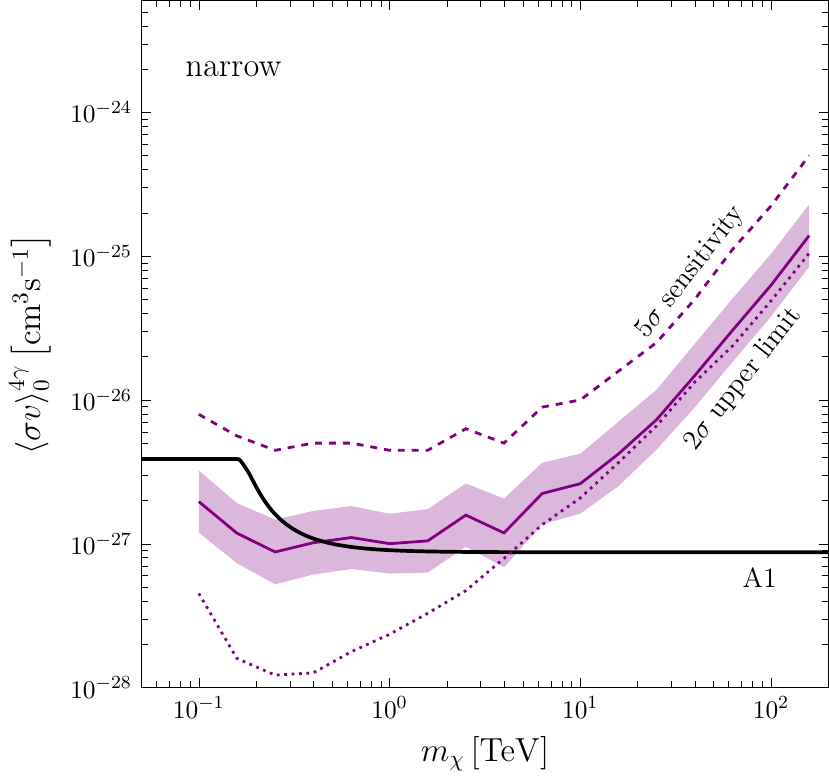}\hspace{-.289cm}
\includegraphics[height=.4\textwidth]{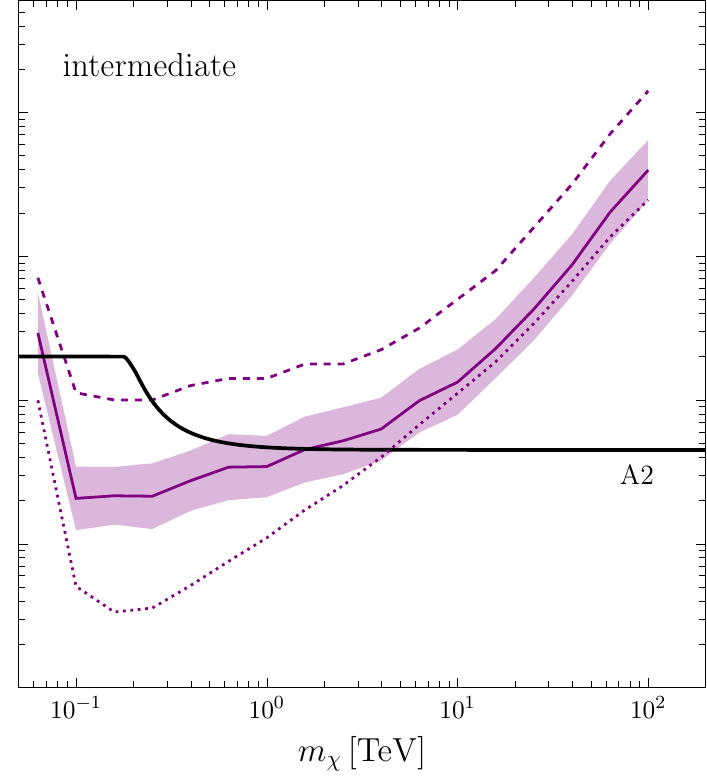}\hspace{-.289cm}
\includegraphics[height=.4\textwidth]{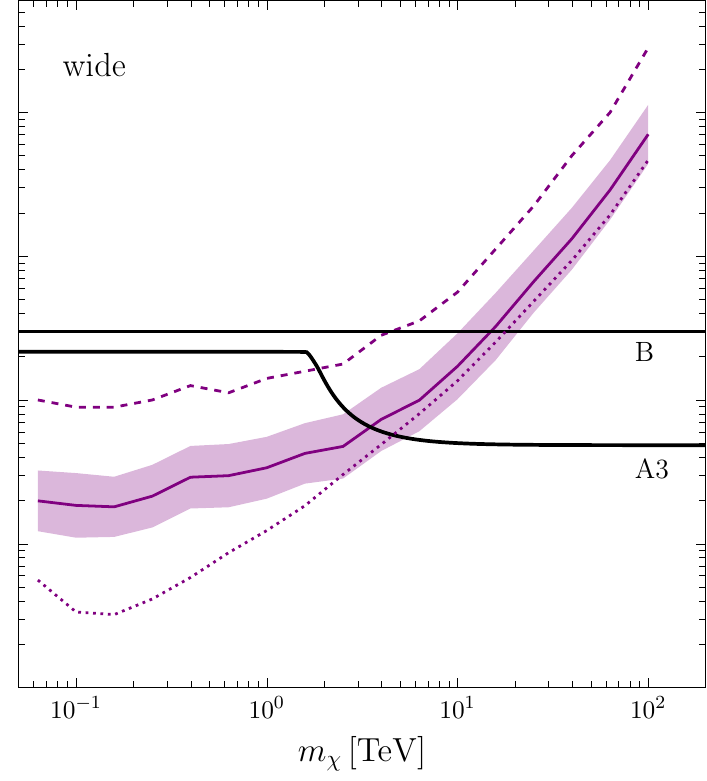}
\hspace{-.2\textwidth}
\caption{\small Expected CTA upper limits and sensitivity reach for gamma-ray boxes in a $2^\circ \times 2^\circ$ region around the Galactic centre after $100\,$h of observations. The left, centre and right panels refer to narrow, intermediate and wide boxes, respectively. The violet solid lines and corresponding bands show the average and standard deviation of the $95\%$ CL upper limits on the annihilation cross section over 300 mock data sets based on the CTA performance of Ref.~\cite{Bernlohr:2012we}, while the $5\sigma$ sensitivities are indicated by the dashed lines. The dotted lines correspond to the average $95\%$ CL upper limits obtained with the updated CTA performance \cite{ctasiteperformance}. Here we assumed an Einasto dark matter profile, a vanishing gamma-ray optical depth and a window width $\epsilon_{2}$. The thick solid lines show the target cross sections corresponding to thermal relics in models A1-3 and B of Tab.~\ref{tab:annbr}.
 }
\label{fig:limits}
\end{center}
\end{figure}

\par The typology of box spectra is rather rich, comprising line-like features at half the dark matter mass, flat shoulders up to the dark matter mass and every variant in-between these two extremes. As discussed in Sec.~\ref{sec:boxes}, the shape of the spectrum is solely defined by the mass ratio $m_\phi/m_\chi$. Here we focus on the three benchmark values $m_\phi/m_\chi=0.999,\,0.9,\,0.1$ that give rise to narrow, intermediate and wide boxes (cf.~right panel of Fig.~\ref{fig:bkg}). Fig.~\ref{fig:limits} shows for each box case the $95\%$ CL upper limits and $5\sigma$ sensitivity reach expected after $100\,$h of CTA observations towards a $2^\circ \times 2^\circ$ region around the Galactic centre. For this figure we have used our baseline assumptions, namely the Einasto profile, a vanishing gamma-ray optical depth and the sliding window of width $\epsilon_{2}$. Generically speaking, the average upper limits cover annihilation cross sections of a few times $10^{-27}\,\text{cm}^3\text{/s}$ or smaller for dark matter masses from about $100\,$GeV up to $10-20\,$TeV, while the expected statistical variance on these limits corresponds to a factor $2-3$. The $5\sigma$ sensitivity reach instead is worse than the average 95\% CL upper limits by a factor $3-5$ across the full dark matter mass range.  Note that the expected prospects for narrow, intermediate and wide boxes are very similar at face value. As noticed in Refs.~\cite{Ibarra:2012dw,Ibarra:2013eda}, this happens because, although narrow features are the most convenient to search for against a smooth background, wider spectra extend to higher energies where the background is dimmer, partially compensating for their relative softness. In Fig.~\ref{fig:limits}, we also show the average $95\%$ CL upper limits obtained with the updated CTA performance \cite{ctasiteperformance}, which are better than the limits based on Ref.~\cite{Bernlohr:2012we} by a factor $\sim8$ (5.6) at low masses for narrow (wide) boxes and at most by a factor of $\sim1.6$ above $4\,$TeV  for all three box sizes. A full study of the implications of the updated CTA performance \cite{ctasiteperformance} is deferred for future work.

\begin{figure}[tp]
\begin{center}
\hspace{-.25\textwidth}
\includegraphics[height=.4\textwidth]{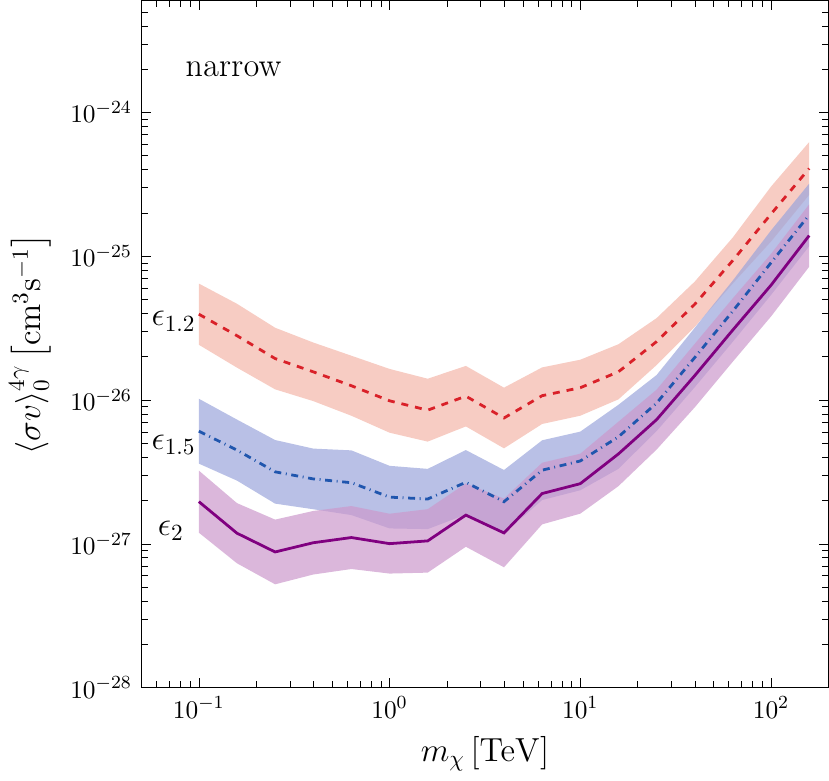}\hspace{-.289cm}
\includegraphics[height=.4\textwidth]{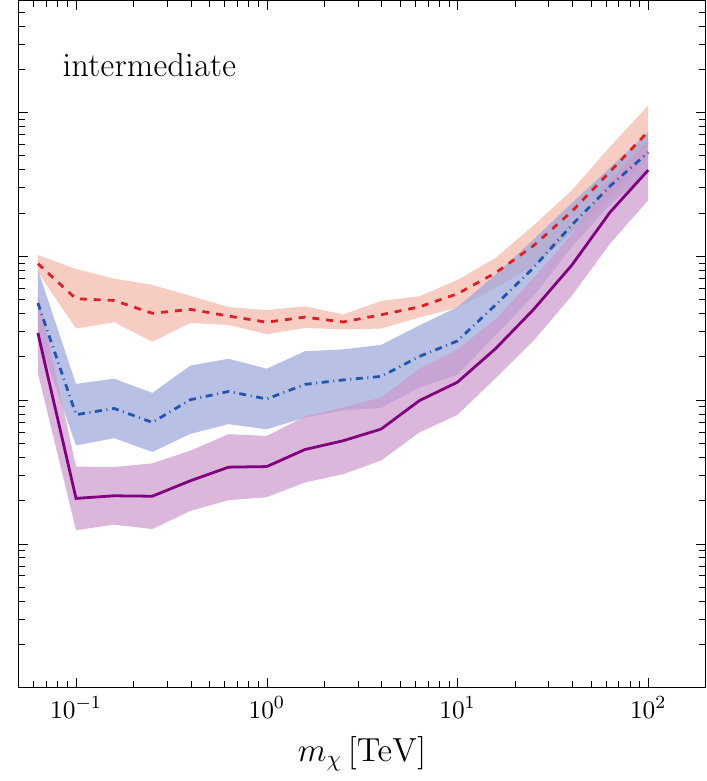}\hspace{-.289cm}
\includegraphics[height=.4\textwidth]{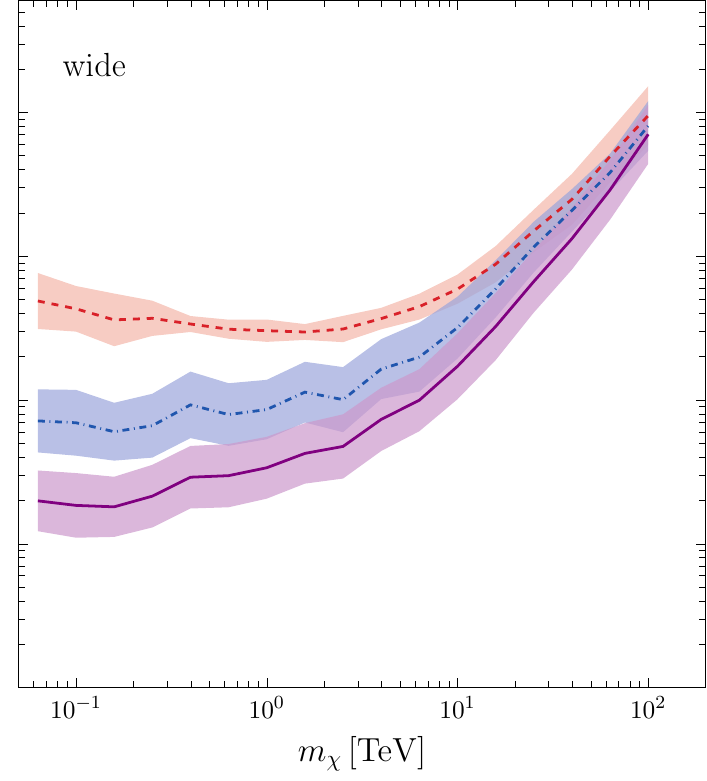}
\hspace{-.2\textwidth}
\caption{\small Expected CTA upper limits for narrow (left panel), intermediate (centre) and wide (right) boxes. In each panel we show the results for the three sliding window widths $\epsilon_{1.2}, \epsilon_{1.5}$ and $\epsilon_{2}$. All unspecified parameters or assumptions are fixed to the ones adopted in Fig.~\ref{fig:limits} and the CTA performance of Ref.~\cite{Bernlohr:2012we} is used.}
\label{fig:window}
\end{center}
\end{figure}

\par We now proceed to check the robustness of the results against our baseline assumptions. The effect of the sliding window on the expected CTA upper limits is presented in Fig.~\ref{fig:window}. Clearly, a wise choice of the window will be crucial to make the best of the available gamma-ray data. This is important for intermediate and wide boxes, where the tightest window of width $\epsilon_{1.2}$ fails to capture a significant part of the signal giving thus the weakest constraints. Such behaviour is also seen for narrow boxes at low energies since there the CTA energy resolution is poor and the expected signal is somewhat extended in energy. In all cases in Fig.~\ref{fig:window}, larger windows can improve the limits as much as a factor 25 provided the background is well described by a power law. We have further checked that our baseline upper limits do not change significantly for different choices of the number of bins $N_b$, and they can be improved up to a factor $1.5-2$ by shifting the window centre $\bar{E}$.

\begin{figure}[tp]
\begin{center}
\hspace{-.25\textwidth}
\includegraphics[height=.4\textwidth]{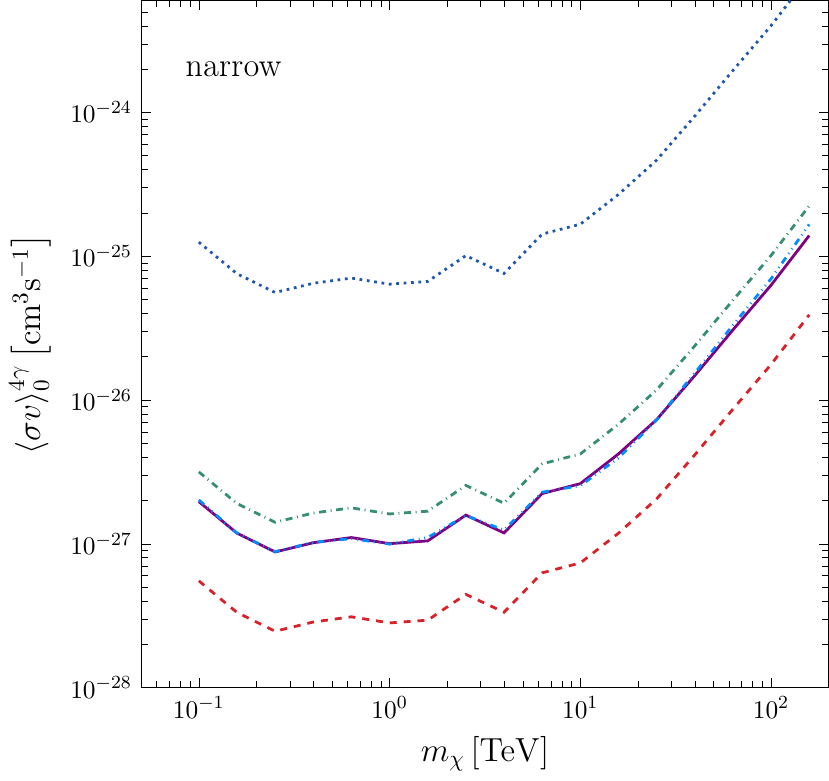}\hspace{-.289cm}
\includegraphics[height=.4\textwidth]{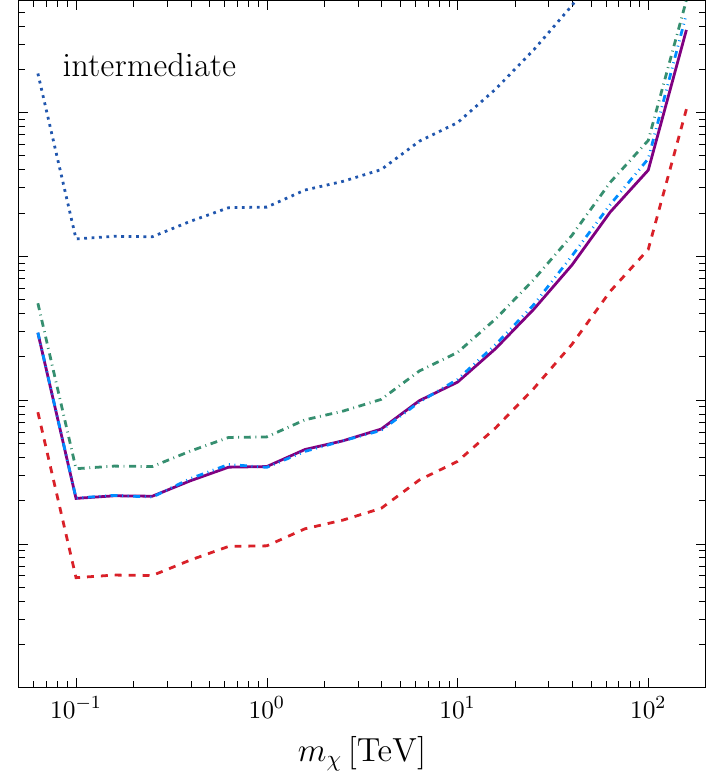}\hspace{-.289cm}
\includegraphics[height=.4\textwidth]{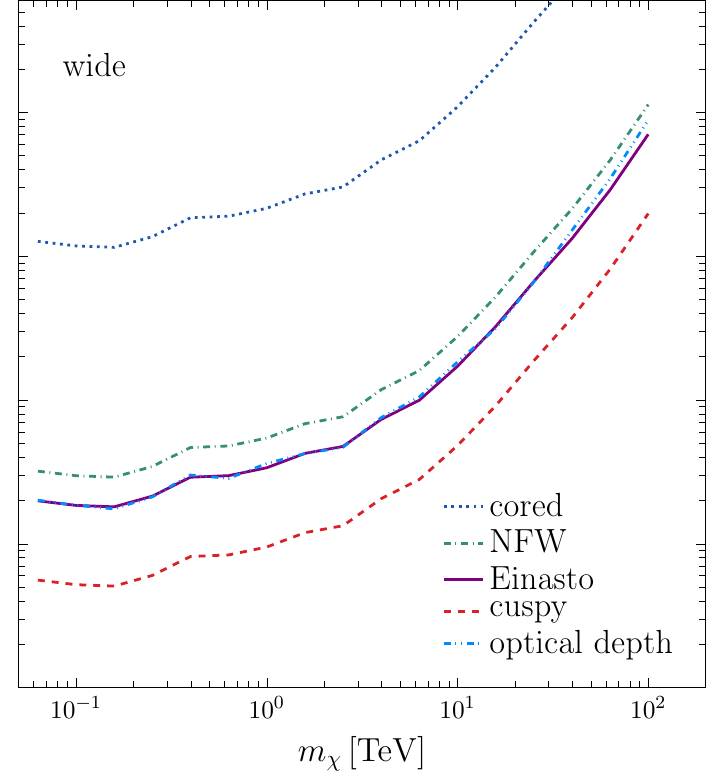}
\hspace{-.2\textwidth}
\caption{\small  Expected CTA upper limits for different dark matter halo profiles, and effect of including the optical depth, for narrow (left panel), intermediate (centre) and wide (right) boxes. In each case we show only the average of the $95\%$ CL upper limits. All unspecified parameters or assumptions are fixed to the ones adopted in Fig.~\ref{fig:limits} and the CTA performance of Ref.~\cite{Bernlohr:2012we} is used.}
\label{fig:profile}
\end{center}
\end{figure}

\par In Fig.~\ref{fig:profile} we show the average $95\%$ CL upper limits obtained for different dark matter profiles. If the dark matter distribution in the inner Galaxy follows a cuspy profile with inner slope $\gamma=1.2$, then our baseline limits reported in Fig.~\ref{fig:limits} improve by roughly a factor $4$, whereas a cored profile leads to a weakening of almost two orders of magnitude. Also shown in Fig.~\ref{fig:profile} is the effect of the gamma-ray optical depth which, according to the estimates at the end of Sec.~\ref{sec:boxpheno}, accounts for an absorption of $25\%$ of the $100\,$TeV photons from dark matter annihilation towards the Galactic centre and deteriorates the limits by around 20\% at the highest masses considered.

\begin{figure}[htp]
\begin{center}
\includegraphics[width=.49\textwidth]{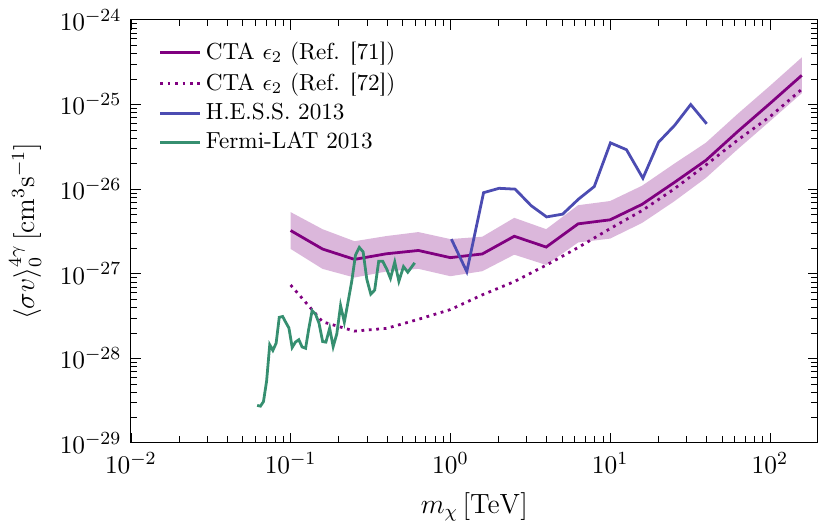}
\includegraphics[width=.49\textwidth]{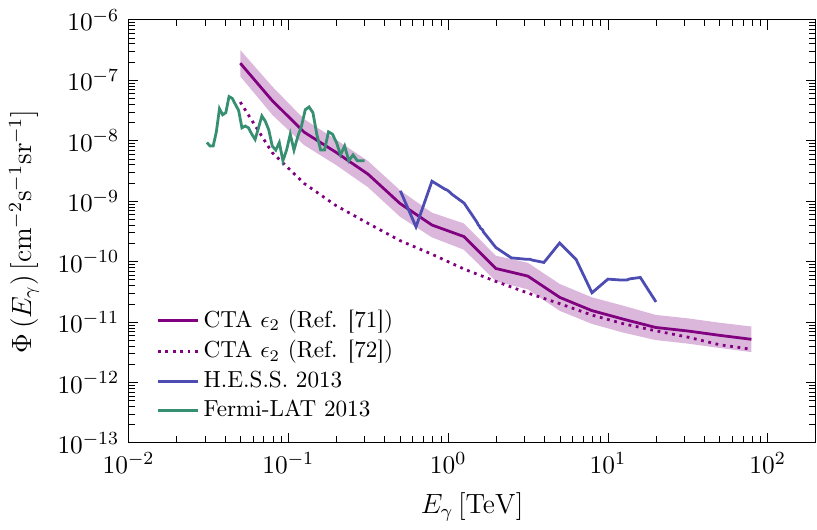}
\caption{\small Comparison of CTA expected upper limits for narrow boxes with current Fermi-LAT \cite{Ackermann:2013uma} and H.E.S.S.~\cite{Abramowski:2013ax} line searches. The left panel shows the upper limits on the cross section into four photons (both Fermi-LAT and H.E.S.S.~results were appropriately rescaled to be compared to narrow boxes, i.e.~lines at half the dark matter mass with four photons per annihilation), while the right panel displays the flux upper limits for a monochromatic signal generated at a given energy. The Fermi-LAT limits correspond to $3.7\,$yr of data towards a circular $3^\circ$ region around the Galactic centre with $J_\text{ann}=1.39\times 10^{23}\,$GeV$^2$\,sr/cm$^5$ for an NFW contracted profile, whereas the H.E.S.S.~searches use $112\,$h observations towards a circular $1^\circ$ region around the Galactic centre excluding $|b|\leq 0.3^\circ$ with $J_\text{ann}=4.43\times 10^{21}\,$GeV$^2$\,sr/cm$^5$ for an Einasto profile.  For the CTA limits shown here we assumed the same observation time, profile, region of interest and J-factor as in H.E.S.S.~\cite{Abramowski:2013ax}, and all other parameters or assumptions are fixed to the ones adopted in Fig.~\ref{fig:limits}. As in Fig.~\ref{fig:limits}, the violet solid lines and bands correspond to the average upper limits obtained with the CTA performance of Ref.~\cite{Bernlohr:2012we}, while the dotted lines are for the updated performance \cite{ctasiteperformance}. All bounds in this figure are $95\%$ CL upper limits.
}
\label{fig:HessFermi}
\end{center}
\end{figure}

\par It is instructive to compare our CTA prospects with the current limits on gamma-ray spectral features. Such comparison is done in the left panel of Fig.~\ref{fig:HessFermi} using the latest line searches from Fermi-LAT \cite{Ackermann:2013uma} and H.E.S.S.~\cite{Abramowski:2013ax}, which have been rescaled from the usual monochromatic line at $m_\chi$ with two photons per annihilation to a narrow box centred at $m_\chi/2$ with four photons per annihilation. The plotted limits correspond to reasonably similar setups, namely a circular $3^\circ$ region around the Galactic centre and an NFW contracted profile for $3.7\,$yr of Fermi-LAT data, and a circular $1^\circ$ region around the Galactic centre excluding $|b|\leq 0.3^\circ$ and an Einasto profile for $112\,$h of H.E.S.S. observations. For the purposes of comparison, we have run a brand new set of CTA constraints matching the observation time, profile, region of interest and J-factor of the H.E.S.S.~limits. The resulting upper limits are overplotted in Fig.~\ref{fig:HessFermi} for the window with width $\epsilon_2$ using both the CTA performance of Ref.~\cite{Bernlohr:2012we} and the updated one \cite{ctasiteperformance}. Our analysis suggests that at energies below a few hundred GeV CTA will hardly supersede the current bounds provided by Fermi-LAT, whereas above $1\,$TeV it will improve upon H.E.S.S.~by up to one order of magnitude.

\par In principle, we could compare our results for narrow boxes to previous works deriving line prospects for CTA, e.g.~\cite{Bringmann:2011ye,Bergstrom:2012vd}. However, let us point out that these works precede the latest CTA design studies of Ref.~\cite{Bernlohr:2012we} (and \cite{ctasiteperformance}), used in our analysis. A direct comparison would therefore be somewhat misleading since the CTA performance (in particular, effective area) implemented in Refs.~\cite{Bringmann:2011ye,Bergstrom:2012vd} differs substantially from Ref.~\cite{Bernlohr:2012we} (and \cite{ctasiteperformance}). Nevertheless, in order to be as transparent as possible and enable a straightforward comparison between different works in the literature, we provide in the right panel of Fig.~\ref{fig:HessFermi} the flux upper limits for monochromatic features at given energy $E_\gamma$, i.e.~$E_\gamma =m_\chi$ for a line or $E_\gamma=m_\chi/2$ for a narrow box.

\par At this point we should comment on the validity and limitations of our results. All the upper limits shown are naturally based on a background-only hypothesis, so our cross section constraints scale as $1/\sqrt{\Delta t}$. We have explicitly verified that this behaviour roughly holds true with an additional run adopting 50$\,$h of CTA observation time. In general, the results presented here are certainly not the last word in terms of CTA prospects: they can be improved with an optimised morphological analysis as pursued e.g.~in Refs.~\cite{2014JCAP...06..024P,2014arXiv1408.4131S,Lefranc:2015pza} and they are limited by the (unknown) systematic level of the instrument as studied in Refs.~\cite{2014arXiv1408.4131S,Lefranc:2015pza}. Following the same approach as in Refs.~\cite{2014arXiv1408.4131S,Lefranc:2015pza}, we find that the effect of systematics in our limits is practically negligible at high masses and sizeable at low masses, in agreement with those references. In any case, the prospects presented in Figs.~\ref{fig:limits} through \ref{fig:HessFermi} constitute the first step to probe gamma-ray boxes from dark matter annihilation over the coming decade.

\begin{figure}[htp]
\includegraphics[width=.8\textwidth]{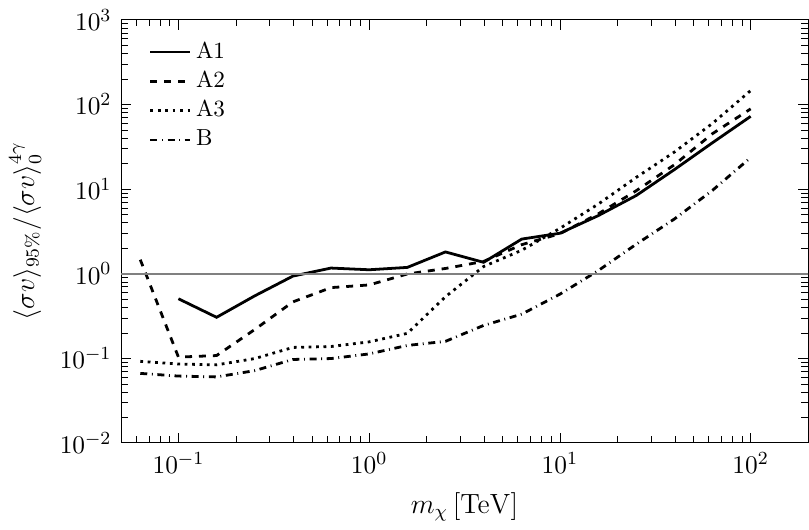}
\begin{center}
\caption{\small The ratio between the average $95\%$ CL upper limit expected for CTA and the target cross section of a thermal dark matter candidate in models A1-3 and B of Tab.~\ref{tab:annbr}. All unspecified parameters or assumptions are fixed to the ones adopted in Fig.~\ref{fig:limits} and the CTA performance of Ref.~\cite{Bernlohr:2012we} is used.} 
\label{fig:ratio}
\end{center}
\end{figure}

\par Finally, we turn to the discussion of the concrete dark matter models introduced in Sec.~\ref{sec:boxmodels}. Let us stress that the use of these models simply serves as illustration for the reach of CTA and is not intended as a comprehensive model building analysis. In Fig.~\ref{fig:limits} we show the cross sections $\langle \sigma v \rangle_{0}^{4\gamma}$ expected from the annihilation of the thermal candidates A1-3 and B in Tab.~\ref{tab:annbr}. For the models A1-3 there is a step-wise behaviour, because the branching ratio BR$(a\to \gamma\gamma)$ falls sharply for $m_a\gtrsim 2m_W$ (cf.~Sec.~\ref{sec:boxmodels}) and because for fixed mass ratio $m_a$ is proportional to $m_\chi$. In the case of models A1 and A2, corresponding to narrow and intermediate boxes, respectively, the sensitivity of CTA is sufficient to probe thermal candidates for relatively low masses (below 1 TeV), a regime in which other instruments (e.g.~Fermi-LAT) provide stronger constraints. The most promising case involves wide boxes as in models A3 and B, for which thermal relics can be tested up to several TeV. We recall that all the above is valid for the assumed (smooth) dark matter profile and concrete particle physics model; therefore for convenience of model builders we provide in Fig.~\ref{fig:ratio} the results of Fig.~\ref{fig:limits} recast in terms of the ratio of the average $95\%$ CL upper limit to the cross section expected from a thermal relic. If we have underestimated the expected signal flux (due to the clumpiness of the dark matter distribution \cite{2008Natur.454..735D,2008MNRAS.391.1685S}, Sommerfeld enhancement \cite{Hisano:2003ec,Hisano:2004ds,Hisano:2006nn} or other) by a modest factor $\sim 10$, then the models A1-3 and B can be probed with CTA up to tens of TeV.

\section{Conclusion}
\label{sec:conclusions}

\par Gamma-ray spectral features at high energies provide perhaps the best way to eventually detect dark matter in indirect searches. Until recently, most efforts in this respect focussed on the search for monochromatic lines and internal bremsstrahlung, both proposed long ago. Gamma-ray boxes constitute a third type of spectral features which opens the possibility to test a whole new class of thermal dark matter candidates. This paper is a first attempt at exploring the capability of future Cherenkov instruments such as CTA to detect or rule out gamma-ray boxes in the high-energy regime. We find that CTA has indeed fairly good discovery prospects for thermal relics in the multi-TeV range, especially so for models featuring wide boxes. In case no signal is found, a large portion of thermal dark matter models will be robustly ruled out up to tens of TeV. This is a golden opportunity for CTA since virtually no other instrument or technique can access this mass regime. The mass range above a few tens of TeV and up to the roof given by unitarity ($m_{\text{dm}}\leq \unit[84]{TeV}$ \cite{Griest:1989wd,Planck:2015xua} for Dirac dark matter) is very challenging to probe even with gamma-ray features and will probably remain open for a very long time.

\vspace{0.5cm}
\section*{Acknowledgements}
We would like to thank David Berge, Christian Farnier, Paolo Panci and Christoph Weniger for useful discussions on the performance of CTA. This research was supported by the DFG cluster of excellence ``Origin and Structure of the Universe'' and by the Graduiertenkolleg ``Particle Physics at the Energy Frontier of New Phenomena''. M.~P.~acknowledges the support from Wenner-Gren Stiftelserna in Stockholm. G.B. acknowledges the support of the European Research Council through the ERC Starting Grant `WIMPs Kairos'.


\bibliographystyle{apsrev.bst}
\bibliography{CTAboxes}

\newpage

{\bf Erratum: On the sensitivity of CTA to gamma-ray boxes from multi-TeV dark matter}

\vspace{0.5cm}
{\bf Abstract} \hspace{0.25cm} Following the erratum of Ref.~\cite{Ibarra:2013eda}, we update here the target cross sections corresponding to the particle physics models A1--3 presented in the published manuscript.

\vspace{0.5cm}

\par The corrections presented in the erratum of Ref.~\cite{Ibarra:2013eda} have an impact in the cross sections and branching ratios of the three benchmark models A1--3 discussed in Sec.~2.2. This update affects Tab.~\ref{tab:annbr} and Figs.~\ref{fig:limits} and \ref{fig:ratio}. In Tab.~\ref{tab:annbr}, after the corrections, all models A1--3 feature $\text{BR}(\chi\bar{\chi} \to as)=1$, while the cross section ratio $\langle \sigma v \rangle_{0}/\langle \sigma v \rangle_{\text{th}}$ is 0.24, 1.02 and 1.01 for models A1, A2 and A3, respectively. Overall, this translates into a target cross section $\langle \sigma v \rangle_{0}^{4\gamma}$ (cf.~Eq.~(2.10)) {\it larger} by a factor of 1.82, 1.51 and 1.39 for models A1, A2 and A3, respectively. Correspondingly, the thick solid lines labelled A1--3 in Fig.~\ref{fig:limits} are to be rescaled upwards by these factors and the lines labelled A1--3 in Fig.~\ref{fig:ratio} are to be rescaled downwards by the same factors. In the process of spotting this issue, we have also found out that in the published version of the manuscript the plots of Fig.~\ref{fig:limits} are missing the benchmark model cross sections (thick solid lines). Therefore, for completeness, we provide in Figs.~\ref{fig:limitsE} and \ref{fig:ratioE} the updated and corrected version of Figs.~\ref{fig:limits} and \ref{fig:ratio}. These changes make the CTA prospects to constrain models A1--3 slightly more optimistic than presented in the published version of the work. None of our conclusions changes.

\renewcommand{\thefigure}{2$'$}

\begin{figure}[h]
\begin{center}
\hspace{-.25\textwidth}
\includegraphics[height=.4\textwidth]{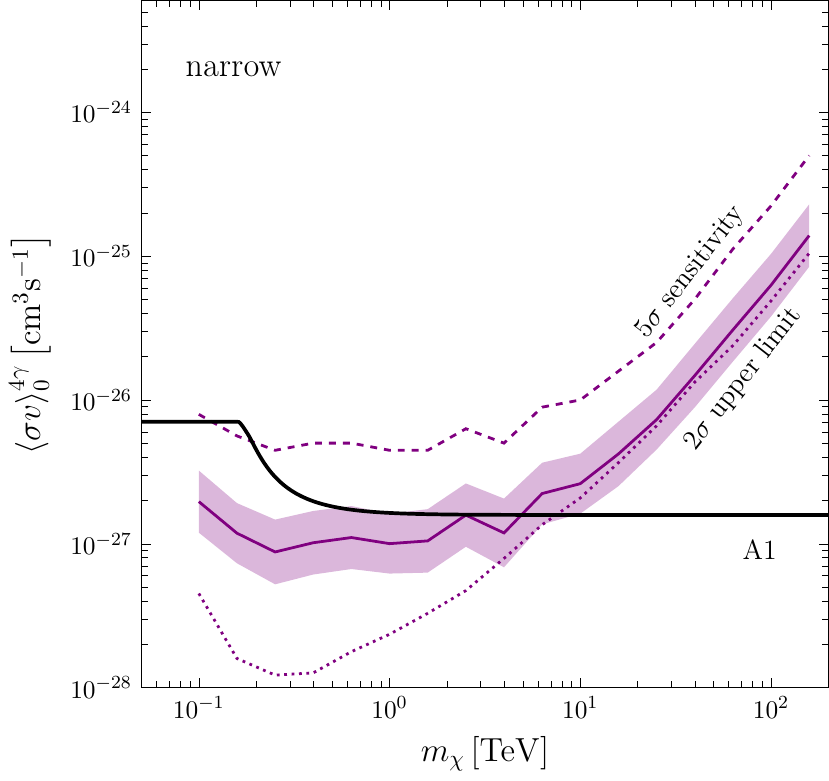}\hspace{-.289cm}
\includegraphics[height=.4\textwidth]{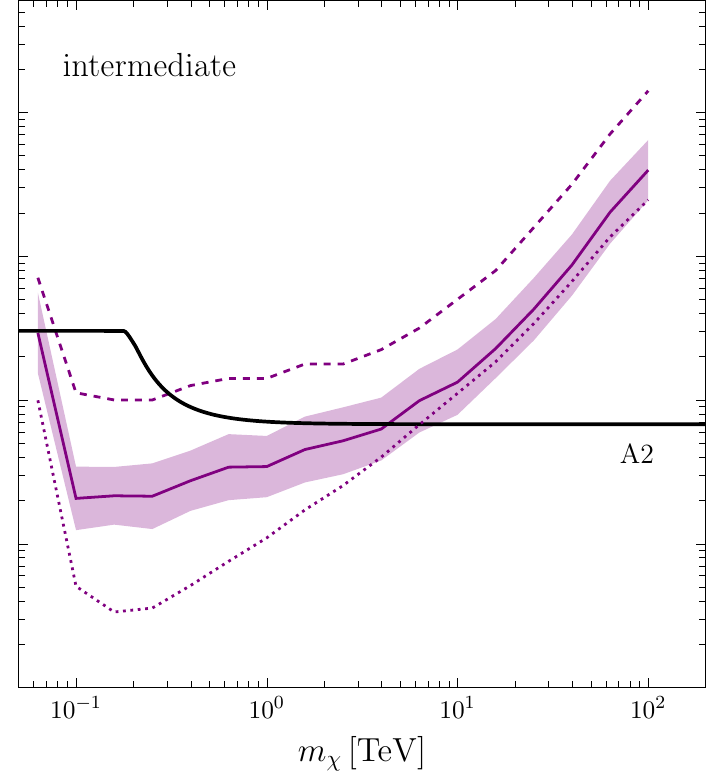}\hspace{-.289cm}
\includegraphics[height=.4\textwidth]{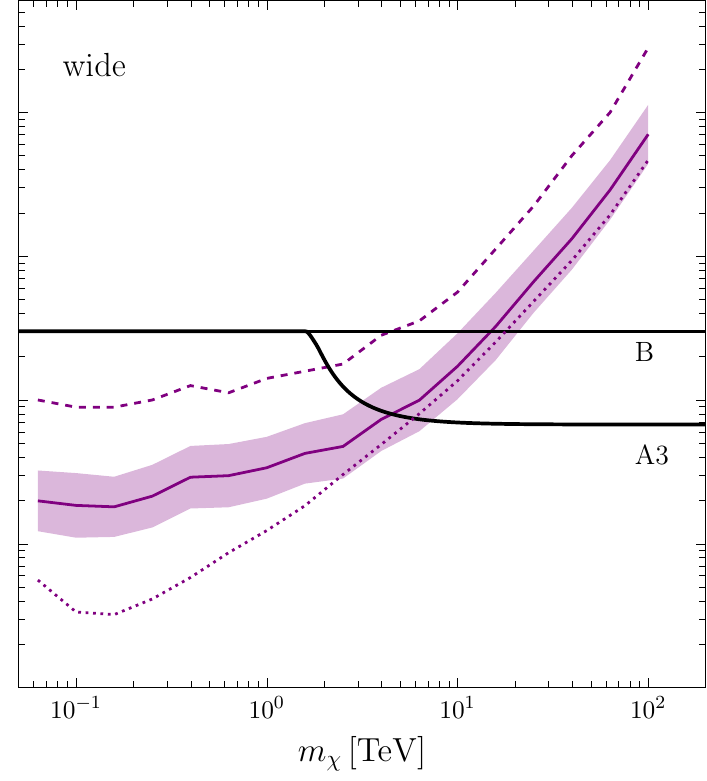}
\hspace{-.2\textwidth}
\caption{\small Expected CTA upper limits and sensitivity reach for gamma-ray boxes in a $2^\circ \times 2^\circ$ region around the Galactic centre after $100\,$h of observations. For a full description, please refer to the caption of Fig.~2 in the published version of the manuscript. Note that here the benchmark cross sections corresponding to models A1--3 have been updated according to the corrections described in the current erratum.
 }
\label{fig:limitsE}
\end{center}
\end{figure}

\renewcommand{\thefigure}{6$'$}

\begin{figure}[htp]
\includegraphics[width=.8\textwidth]{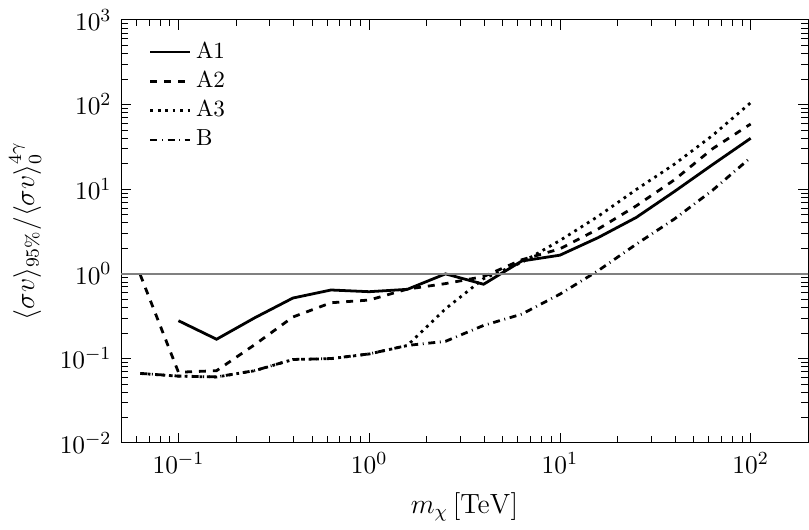}
\begin{center}
\caption{\small The ratio between the average $95\%$ CL upper limit expected for CTA and the target cross section of a thermal dark matter candidate in models A1--3 and B of Tab.~1. For a full description, please refer to the caption of Fig.~6 in the published version of the manuscript. Note that here the benchmark cross sections corresponding to models A1--3 have been updated according to the corrections described in the current erratum.
} 
\label{fig:ratioE}
\end{center}
\end{figure}

\vspace{0.5cm}
\section*{Acknowledgements}
The authors thank H.M.~Lee and W.-I.~Park for useful discussions.
This research was supported by the DFG cluster of excellence `Origin and Structure of the Universe'. M.~P.~acknowledges the support from Wenner-Gren Stiftelserna in Stockholm.


\end{document}